\documentclass[12pt,a4paper]{article}
\usepackage{amsmath}
\usepackage{amssymb}
\usepackage[nosort]{cite}

\begin{document}

\newcommand{\indup}[1]{_{\mathrm{#1}}}
\newcommand{\hypref}[2]{\ifx\href\asklfhas #2\else\href{#1}{#2}\fi}

\newcommand{\sfrac}[2]{{\textstyle\frac{#1}{#2}}}
\newcommand{\half}{\sfrac{1}{2}}
\newcommand{\quarter}{\sfrac{1}{4}}

\newcommand{\Op}{\mathcal{O}}
\newcommand{\order}[1]{\mathcal{O}(#1)}
\newcommand{\eps}{\varepsilon}
\newcommand{\Lagr}{\mathcal{L}}
\newcommand{\superN}{\mathcal{N}}
\newcommand{\gym}{g_{\scriptscriptstyle\mathrm{YM}}}
\newcommand{\gtwo}{g_2}
\newcommand{\Tr}{\mathop{\mathrm{Tr}}}
\renewcommand{\Re}{\mathop{\mathrm{Re}}}
\renewcommand{\Im}{\mathop{\mathrm{Im}}}
\newcommand{\Li}{\mathop{\mathrm{Li}}\nolimits}
\newcommand{\cdott}{\mathord{\cdot}}
\newcommand{\singlet}{{\mathbf{1}}}

\newcommand{\lrbrk}[1]{\left(#1\right)}
\newcommand{\bigbrk}[1]{\bigl(#1\bigr)}
\newcommand{\vev}[1]{\langle#1\rangle}
\newcommand{\normord}[1]{\mathopen{:}#1\mathclose{:}}
\newcommand{\lrvev}[1]{\left\langle#1\right\rangle}
\newcommand{\bigvev}[1]{\bigl\langle#1\bigr\rangle}
\newcommand{\bigcomm}[2]{\big[#1,#2\big]}
\newcommand{\lrabs}[1]{\left|#1\right|}
\newcommand{\abs}[1]{|#1|}

\newcommand{\nn}{\nonumber}
\newcommand{\nln}{\nonumber\\}
\newcommand{\nl}{\nonumber\\&&\mathord{}}
\newcommand{\nle}{\nonumber\\&=&\mathrel{}}
\newcommand{\eq}{\mathrel{}&=&\mathrel{}}
\newenvironment{myeqnarray}{\arraycolsep0pt\begin{eqnarray}}{\end{eqnarray}\ignorespacesafterend}
\newenvironment{myeqnarray*}{\arraycolsep0pt\begin{eqnarray*}}{\end{eqnarray*}\ignorespacesafterend}

\newcommand{\ft}[2]{{\textstyle\frac{#1}{#2}}}
\newcommand{\cO}{{\cal O}}
\newcommand{\cT}{{\cal T}}
\def\ss{\scriptstyle}
\def\st{\scriptstyle}
\def\sst{\scriptscriptstyle}
\def\ra{\rightarrow}
\def\lra{\longrightarrow}
\newcommand\Zb{\bar Z}
\newcommand\bZ{\bar Z}
\newcommand\bF{\bar \Phi}
\newcommand\bP{\bar \Psi}

\thispagestyle{empty}
\begin{flushright}
{\sc\footnotesize hep-th/0212269}\\
{\sc AEI 2002-104}
\end{flushright}
\vspace{1cm}
\setcounter{footnote}{0}
\begin{center}
{\Large{\bf BMN Gauge Theory as a Quantum Mechanical System\par}
    }\vspace{7mm}
{\sc N. Beisert, C. Kristjansen,  
J. Plefka and M. Staudacher\\[7mm]
Max-Planck-Institut f\"ur Gravitationsphysik\\
Albert-Einstein-Institut\\
Am M\"uhlenberg 1, D-14476 Golm, Germany}\\ [2mm]
{\tt nbeisert,kristjan,plefka,matthias@aei.mpg.de}\\[20mm]

{\sc Abstract}\\[2mm]
\end{center}

We rigorously derive an effective quantum mechanical 
Hamiltonian from ${\cal N}=4$ gauge theory in the BMN limit. 
Its eigenvalues yield the exact one-loop anomalous dimensions 
of scalar two-impurity BMN operators for all genera. It is
demonstrated that this reformulation vastly simplifies 
computations. E.g.~the known anomalous dimension formula 
for genus one is reproduced through a one-line 
calculation. We also efficiently evaluate the
genus two correction, finding a non-vanishing result.
We comment on multi-trace two-impurity operators 
and we conjecture that our quantum-mechanical reformulation 
could be extended to higher quantum loops and more impurities.

\newpage
\setcounter{page}{1}

In~\cite{Berenstein:2002jq} a modified, simpler version of the
AdS/CFT duality between IIB string theory and ${\cal N}=4$ supersymmetric
gauge theory was considered. It involves taking a limit on both sides
of the correspondence. On the string side the Penrose limit of the
AdS$_5 \times $S$^5$ background results in the so-called
plane-wave background~\cite{Blau:2002dy,Metsaev:2001bj,Metsaev:2002re},
while the limit on the gauge theory side leads to the consideration
of operators of large $R$-charge $J$ in conjunction with
a large $N$ limit.
After identification of the correct operators~\cite{Berenstein:2002jq}
it is possible to relate and match, for various massless and massive states, 
the mass spectrum of free plane-wave
string excitations~\cite{Metsaev:2001bj,Metsaev:2002re} to the planar 
scaling dimensions of the corresponding 
Berenstein-Maldacena-Nastase (BMN)
operators\cite{Berenstein:2002jq,Gross:2002su,Santambrogio:2002sb}.

It is natural to study the extension of the BMN correspondence to the case of
interacting strings. On the gauge theory side, this involves
the consideration of corrections to the planar limit.
It was demonstrated in
\cite{Kristjansen:2002bb,Constable:2002hw,Beisert:2002bb,Constable:2002vq}
that such corrections indeed survive the BMN limiting procedure.
As far as the scaling dimensions of BMN operators 
are concerned, the corrections turned out to be finite.
Much work~
\cite{Spradlin:2002ar,Chu:2002pd,Spradlin:2002rv,
Chu:2002qj,Chu:2002eu,Pankiewicz:2002gs,Pankiewicz:2002tg,Chu:2002wj,He:2002zu}
has also been undertaken to develop a string field theory approach, which
should eventually permit to obtain the string corrections to the
free mass formula of plane-wave states.
A recent effort to derive the gauge theory results 
~\cite{Kristjansen:2002bb,Constable:2002hw,Beisert:2002bb,Constable:2002vq}
for genus one 
from string theory has been reported in \cite{Roiban:2002xr}. 
An alternative, more heuristic approach, which is conceptionally and 
methodologically
interpolating between string and gauge theory has been pursued in
refs.~\cite{Verlinde:2002ig,Vaman:2002ka,Pearson:2002zs,Gomis:2002wi}.

In order to unearth its relationship to string theory,
it is important to understand 
the structure of BMN gauge theory as completely as possible.
It is widely suspected (see most of the above references) that
the formalism underlying the limit is essentially one-dimensional
and should therefore resemble the description of a quantum mechanical system.
In this note we shall demonstrate, in the example of two impurities and
on the one-loop level, but to arbitrary order in the topological
expansion, that this is indeed the case. In fact, we will {\it derive}
the parts of the Hamiltonian relevant to our situation directly from
the gauge theory. Aside from providing this conceptual insight, 
we shall also show how the rather laborious computations 
of~\cite{Kristjansen:2002bb,Constable:2002hw,Beisert:2002bb,Constable:2002vq}
can be significantly simplified and extended.
An important first hint of this hidden simplicity was discovered by
Janik~\cite{Janik:2002bd}, and we will complete and considerably
extend his arguments. For steps towards directly deriving a Hamiltonian
from ${\cal N}=4 $ gauge theory by dimensional reduction, see
\cite{Okuyama:2002zn}.  

Recall~\cite{Kristjansen:2002bb,Constable:2002hw,
Beisert:2002bb,Constable:2002vq} that the BMN suggestion leads to a 
double-scaling limit of the ${\cal N}=4$ gauge theory: The $R$-charge
$J$ and the number of colors $N$ tend to infinity such that
the effective quantum loop counting parameter
$\lambda'=\gym^2 N/J^2$ and the effective genus counting parameter
$g_2^2=J^4/N^2$ stay finite. The quantities of physical interest are 
the anomalous dimensions of the BMN operators. 
In~\cite{Constable:2002hw} it was conjectured that {\it physical} quantities
involving string interactions should not depend on
$g_2$ but instead on an {\it effective} string coupling 
constant $g_2~\sqrt{\lambda'}$. 
This also appears to be implicit in the work of
\cite{Verlinde:2002ig,Vaman:2002ka,Pearson:2002zs}.
It would seem to imply that at the 
one-loop (${\cal O}(\lambda'$)) level no corrections from genera higher than
one should be allowed. Our new procedure allows us to easily perform
a genus two calculation of the one-loop anomalous dimensions,
which would have been horrendously complicated
with the previous methods. We will find a non-vanishing result below,
which appears to disprove the just mentioned conjecture.

The scalar BMN operators are made from the three 
complex scalar fields $Z,\phi,\psi$ of the gauge theory. 
They are obtained by modifying charge $J$ BPS ($k+1$)-trace operators
$\Tr Z^{J_0} \Tr Z^{J_1} \ldots \Tr Z^{J_k}$ 
(where $J=J_0 +J_1 +\ldots +J_k$) by doping them with a number 
of ``impurities''\cite{Berenstein:2002jq} $\phi, \psi$:
\begin{equation}
\label{operators}
\cO_p^{J_0,J_1,\ldots,J_k}=\Tr (\phi Z^p\psi Z^{J_0-p}) \, \Tr Z^{J_1}
\ldots\Tr Z^{J_k}
\end{equation}
where $0 \leq p \leq J_0$.
Here we will only consider the simplest case of two impurities.
At the classical level the doped operators have
conformal dimension $J+2$. This  dimension gets corrected once
interactions are turned on: The operators are no longer protected 
if the impurities are inserted into the same trace.
However, 
it is possible to find linear combinations 
of the $\Op_{\alpha}$ ($\alpha$ is a multi-index) which
have a definite conformal dimension. Conformal scaling operators 
are characterized by being eigenstates of the dilatation operator
$D$ with eigenvalue $\Delta$, which equals their conformal dimension. 
In the planar limit it is possible
to consider single trace operators alone and 
the corrections are given by the BMN
prediction~\cite{Berenstein:2002jq}. For a precise definition of
these operators at finite $J$ see~\cite{Beisert:2002tn}.
When one takes into account non-planar contributions 
single-trace operators no longer 
suffice~\cite{Bianchi:2002rw,Beisert:2002bb,Constable:2002vq} and we have to
find the appropriate linear combinations in the class of
multi-trace operators in eq.\eqref{operators}.

The matrix elements of the dilatation operator
in the BMN basis can be read off from the two-point functions of the
BMN operators. With $\alpha$ and $\beta$ multi-indices we can write a
one-loop two-point function of BMN operators as
\begin{equation}
\bigvev{\Op_{\alpha}(x)\, \bar\Op_{\bar\beta}(0)}
=\frac{1}{|x|^{2(J+2)}}\left(S_{\alpha\bar\beta}+T_{\alpha\bar\beta} 
\log(|x\Lambda|^{-2})\right)
\label{twopoint}
\end{equation}
where $\Lambda$ is a (divergent) renormalization constant
(see e.g.~\cite{Beisert:2002bb}).
Here $S_{\alpha\bar\beta}$ is the tree level mixing matrix while 
$T_{\alpha\bar\beta}$ encodes the interactions. As pointed out in 
references~\cite{Gross:2002mh,Janik:2002bd} the 
one-loop dilatation operator
matrix element, $D_{\alpha}{}^{\beta}$,
defined by $D\,\Op_{\alpha}=D_{\alpha}{}^{\beta}\Op_{\beta}$, 
can be expressed as
\begin{equation}
D_{\alpha}{}^\beta=(J+2)\,\delta_{\alpha}{}^{\beta}+
T_{\alpha\bar\gamma}(S^{-1})^{\bar\gamma\beta}.
\label{Dmatrix}
\end{equation}
The matrix elements $S_{\alpha\bar\beta}$ can be identified as
expectation values in a zero-dimensional Gaussian complex matrix
model and efficiently evaluated using matrix model 
techniques~\cite{Kristjansen:2002bb,Eynard:2002df}. 
We write:
\begin{equation}\label{eq:freecorr}
S_{\alpha\bar\beta}=\bigvev{\Op_{\alpha} \,\bar\Op_{\bar\beta}}
\end{equation}
where all fields appearing in $\Op_{\alpha}(x)$ 
have been replaced with space-time independent fields.
At the one-loop level a
similar simplification can be achieved in the case of
$T_{\alpha\bar\beta}$ by making use of an effective vertex inside the matrix
model correlator
which captures the sum of all contributing one-loop Feynman diagrams
\cite{Kristjansen:2002bb,Constable:2002hw}. More
precisely, one has
\begin{equation}\label{eq:oneloopcorr}
T_{\alpha\bar\beta}=\bigvev{\Op_{\alpha}\, H\, \bar\Op_{\bar\beta}}
\end{equation}
where
\begin{equation}
H=-\frac{\gym^2}{8 \pi^2}\,\,\normord{\bigbrk{\Tr [\bar{Z},\bar{\phi}][Z,\phi]
+\Tr [\bar{Z},\bar{\psi}][Z,\psi]+\Tr [\bar{\phi},\bar{\psi}][\phi,\psi]}}
\label{effvertex}
\end{equation}
In the correlator, two of the legs of $H$ connect to $\Op_\alpha$ and the other
two to $\bar \Op_{\bar\beta}$. 
The normal ordering symbol~: :~means that pairwise contractions between 
fields inside $H$ are to be omitted.
The remaining fields are contracted among each other 
by free matrix model correlators. The order in which the contractions are
performed is irrelevant, so one may first contract the
effective vertex with $\Op_\alpha$. At this point one notices that the 
fields which are generated  after the contraction
are again given by a linear combination of the operators $\Op$:
\begin{equation}
H\circ\Op_\alpha=H_\alpha{}^\gamma\, \Op_\gamma.
\end{equation}
By making use of \eqref{eq:freecorr}
the one-loop correlator \eqref{eq:oneloopcorr}
is readily evaluated
\begin{equation}
T_{\alpha\bar\beta}=\bigvev{(H\circ\Op_{\alpha})\, \bar\Op_{\bar\beta}} 
=H_\alpha{}^\gamma \bigvev{\Op_{\gamma}\, \bar\Op_{\bar\beta}} 
=H_\alpha{}^\gamma S_{\gamma\bar\beta}.
\end{equation}
Comparing this to eq.\eqref{Dmatrix} we find
\begin{equation}D_\alpha{}^\beta=(J+2)\,\delta_\alpha{}^\beta+H_\alpha{}^\beta.\end{equation}
We see that the matrix $H_{\alpha}{}^{\beta}$ is the one-loop part of the dilatation
matrix $D_{\alpha}{}^{\beta}$ and conclude that
the two matrices have the same eigenvectors. 
Thus diagonalizing the matrix $H$ will immediately give us the anomalous 
dimension of the BMN operators. 
In particular, we will never need to know the 
explicit form of the tree level mixing matrix $S_{\alpha\beta}$.
Furthermore, the eigenvectors $\hat\Op_{\alpha}$ of $H$ can be identified
with the BMN operators up to normalization constants $C_{\alpha}$.
It is obvious from conformal field theory
that the eigenoperators
of the dilatation operator have orthogonal correlators
\begin{equation}
\bigvev{\hat\Op_{\alpha}\,\hat{\bar\Op}_{\bar\beta}}
=\delta_{\alpha{\bar\beta}}\, |C_{\alpha}|^2.
\end{equation}
In other words they are orthogonal with 
respect to the inner product induced by the
mixing matrix $S_{\alpha\bar\beta}$ at tree-level. 
To obtain the normalization constants $C_{\alpha}$,
however, would require the knowledge of the 
tree-level mixing matrix.

Let us, as described above, consider the action of the effective vertex 
$H$ on the general state eq.\eqref{operators}; performing
the Wick contractions one finds
that $H:=H_0+H_+ + H_-$ with
\begin{myeqnarray}\label{lattice}
{H}_0\circ \cO_p^{J_0,J_1,\ldots,J_k}\eq
-\frac{\gym^2N}{4\pi^2}\left[\cO^{J_0,J_1,\ldots,J_k}_{p+1}
-2\, \cO^{J_0,J_1,\ldots,J_k}_p 
+\,\cO^{J_0,J_1,\ldots,J_k}_{p-1}
\, \right]  ,
\\
{H}_+\circ \cO_p^{J_0,J_1,\ldots,J_k}\eq
\frac{\gym^2}{4\pi^2}\sum_{J_{k+1}=1}^{p-1}\, 
\left( 
\cO^{J_0-J_{k+1},J_1,\ldots,J_{k+1}}_{p-J_{k+1}} -
\cO_{p-1-J_{k+1}}^{J_0-J_{k+1},J_1,\ldots,{J_{k+1}}} 
\right) 
\nl
-\frac{\gym^2}{4\pi^2}\sum_{J_{k+1}=1}^{J_0-p-1}\,
\left( 
\cO^{J_0-J_{k+1},J_1,\ldots,J_{k+1}}_{p+1}
-\cO^{J_0-J_{k+1},J_1,\ldots,J_{k+1}}_p 
\right),
\nln
{H}_-\circ \cO_p^{J_0,J_1,\ldots,J_k}\eq
\frac{\gym^2}{4\pi^2}\sum_{i=1}^k J_i\, 
\left(
\cO^{J_0+J_i,J_1,\ldots,\makebox[0pt]{\,\,\,\,$\times$}J_{i},\ldots,J_k}_{J_i+p} 
-\cO^{J_0+J_i,J_1,\ldots,\makebox[0pt]{\,\,\,\,$\times$}J_{i},\ldots,J_k}_{J_i+p-1}
\right) 
\nl
-\frac{\gym^2}{4\pi^2} \sum_{i=1}^k J_i\, \left( 
\cO^{J_0+J_i,J_1,\ldots,\makebox[0pt]{\,\,\,\,$\times$}J_{i},\ldots,J_k}_{p+1}
-\cO^{J_0+J_i,J_1,\ldots,\makebox[0pt]{\,\,\,\,$\times$}J_{i},\ldots,J_k}_p
\right). \nn
\end{myeqnarray}
Here we have neglected boundary terms and
interactions involving both impurities at the same time. 
In the BMN limit these will not contribute.
The states $\cO_p^{J_0,J_1,\ldots,J_k}$ are not orthogonal in the
gauge theory, but
this is not necessary for
finding the spectrum of $H$. 
We observe that
the special multi-trace states $\cO_p^{J_0,J_1,\ldots,J_k}$ form a 
complete set as far as the action of the operator $H$ is
concerned. $H_0$ is trace-number conserving, and 
$H_+$ and $H_-$ respectively increase and decrease
the number of traces by one. (Clearly we have $H_- \circ \cO^{J_0}_p=0$).
Interestingly, we see that the set of BPS type two-impurity
operators $\Tr (\phi Z^{J_1})  \Tr ( \psi Z^{J_2}) \Tr Z^{J_3} \ldots$
completely decouples. We can
now imagine the $J_i$ to be very large so that we can view
$x:=p/J$ and $r_i:=J_i/J$ as continuous variables, allowing us
to formulate the spectral problem directly in the BMN limit.
We therefore replace the discrete states eq.\eqref{operators}
by a set of continuum states
\begin{equation}
\label{states}
\cO_p^{J_0,J_1,\ldots,J_k} \rightarrow |x;r_1, \ldots, r_k \rangle
\end{equation}
spanning a Hilbert space, where
\begin{equation}
x \in [0,r_0],\qquad
r_0,r_i\in [0,1] \qquad \mbox{and}\qquad r_0=1-(r_1+ \ldots +r_k).
\end{equation}
Clearly the states eq.\eqref{states} are invariant under arbitrary
permutations $\pi \in \;$S$_k$ of the trace numbers:
$|x;r_1, \ldots, r_k \rangle = |x;r_{\pi(1)}, \ldots, r_{\pi(k)} \rangle$.
Now it is easy to derive the continuum limit of eqs.\eqref{lattice}.
Defining
\begin{equation}
\label{h}
H=\frac{\lambda'}{4\pi^2}\,h, \qquad \lambda'=\frac{\gym^2 N}{J^2}\qquad\mbox{and}\qquad \gtwo=\frac{J^2}{N}
\end{equation}
the action of $h:=h_0+\gtwo\,h_+ +\gtwo\,h_-$ 
on the continuum states in eq.\eqref{states} can be written as
\begin{myeqnarray}\label{haction}
h_0\, |x;r_1, \ldots, r_k \rangle \eq -\partial_x^2~
|x;r_1, \ldots, r_k \rangle  ,
\\\nn
\\
h_+\, |x;r_1, \ldots, r_k \rangle \eq
\int_0^x dr_{k+1}\, ~\partial_x 
~|x-r_{k+1};r_1, \ldots, r_{k+1} \rangle
\nl
-\int_0^{r_0-x}dr_{k+1}\, 
~\partial_x 
~|x;r_1, \ldots, r_{k+1} \rangle ,
\nln
h_-\, |x;r_1, \ldots, r_k \rangle \eq
\sum_{i=1}^k ~r_i~\partial_x~
|x+r_i;r_1, \ldots, \makebox[0pt]{\,\,$\times$}r_{i},\ldots,r_k \rangle
\nl
-\sum_{i=1}^k ~r_i~\partial_x~
|x;r_1, \ldots, \makebox[0pt]{\,\,$\times$}r_{i},\ldots,r_k \rangle.
\nn
\end{myeqnarray}
We note that this Hamiltonian manifestly terminates at $\order{\gtwo}$. 
There are no contact terms.

We will now first diagonalize our states w.r.t.~the trace-number
conserving, free Hamiltonian $h_0$, which is exact iff $g_2=0$. 
The $(k+1)$-trace eigenstates (with $n$ integer) are
\begin{equation}
\label{fourier}
|n;r_1, \ldots, r_k \rangle = \frac{1}{\sqrt{r_0}}\,
\int_0^{r_0} dx\, e^{2\pi i n x/r_0}\, 
|x;r_1, \ldots, r_k \rangle,
\end{equation}
they obey 
\begin{equation}
\label{h0action}
h_0\, |n;r_1,\ldots, r_k\rangle =
E^{(0)}_{|n;r_1,\ldots,r_k\rangle} |n;r_1,\ldots, r_k\rangle
\end{equation}
with ``energy'' eigenvalues (i.e. anomalous dimensions)
\begin{equation}
\label{freespectrum}
E^{(0)}_{|n;r_1,\ldots,r_k\rangle}=4 \pi^2~\frac{n^2}{r_0^2}.
\end{equation}
For multi-trace states the spectrum is continuous, while single-trace
states (where $k=0$ and $r_0=1$), corresponding to the original
BMN operators~\cite{Berenstein:2002jq}, have a purely discrete spectrum.
Now we can proceed to evaluate the topological corrections to the
energies by standard quantum mechanical
perturbation theory. 

Let us evaluate the action of the interaction piece 
$h'=h_+ + h_-$ of our Hamiltonian on the free eigenstates. 
{}From eqs.\eqref{haction},\eqref{fourier} we find for the 
trace-creation and trace-annihilation operators
\begin{myeqnarray}
\label{h-on-eigen}
\lefteqn{
h_+\,  |n;r_1, \ldots, r_k \rangle =\nn }\\  
&&\qquad\int_0^{r_0} dr_{k+1}\, \sum_{m=-\infty}^{\infty}
\,\,\frac{4 m \sin^2\bigbrk{\pi n\sfrac{r_{k+1}}{r_0}}}
{\sqrt{r_0}\sqrt{r_0-r_{k+1}}\bigbrk{m-n \frac{r_0-r_{k+1}}{r_0}}}
~|m;r_1, \ldots, r_{k+1} \rangle ,
\nn \\
\lefteqn{ h_-\,  |n;r_1, \ldots, r_k \rangle = \nn}\\ 
&&\qquad\sum_{i=1}^k\; \sum_{m=-\infty}^{\infty}
\,\,\frac{4\, r_i\, m \sin^2\bigbrk{\pi m\sfrac{r_i}{r_0+r_i}}}
{\sqrt{r_0}\sqrt{r_0+r_i}\bigbrk{m-n \frac{r_0+r_i}{r_0}}}
~|m;r_1, \ldots, \makebox[0pt]{\,\,$\times$}r_{i},\ldots,r_k \rangle .
\end{myeqnarray} 
Note that $r_0=1-(r_1+\ldots+r_k)$ is defined to be the size of the first trace
of the operator on the left-hand side of the equations.

The one-loop anomalous dimensions of the BMN operators 
equal, up to a factor of $\lambda'/4\pi^2$,
the eigenvalues of the operator $h$:
\begin{equation}\label{eq:eigenaction}
h\,|\hat n;r_1,\ldots, r_k\rangle=
E_{|n;r_1,\ldots, r_k\rangle}\,|\hat n;r_1,\ldots, r_k\rangle.
\end{equation}
As discussed above, the exact eigenstates 
$|\hat n;r_1,\ldots, r_k\rangle$ correspond, up to normalization,
to the diagonalized BMN operators. 
At $\gtwo=0$ we have already diagonalized the Hamiltonian,
therefore we can proceed to evaluate the energies 
as an expansion in the genus counting parameter.
As our perturbation $h'$ is entirely off-diagonal, energies 
determined by non-degenerate perturbation theory
will be given
as a series in the square of the perturbation parameter $\gtwo$ in accordance
with the nature of the gauge theory genus expansion
\begin{equation}\label{eq:eigenenergies}
E_{|n;r_1,\ldots, r_k\rangle}=
\sum_{h=0}^\infty \,\gtwo^{2h}\,E^{(h)}_{|n;r_1,\ldots, r_k\rangle}.
\end{equation}
Correspondingly, the exact eigenstates $|\hat n;r_1,\ldots,r_k\rangle$
are linear combinations of the bare states $|m;s_1,\ldots,s_l\rangle$.
The mixing coefficients are power series in $\gtwo$.
The free Hamiltonian $h_0$ immediately gives us 
the energies at the spherical level, \emph{cf.}~\eqref{h0action}.
Higher genera contributions can be obtained by
quantum mechanical perturbation theory. 

At this point it seems convenient to introduce a
scalar product on the space of states.
\begin{equation}
\langle n;s_1,\ldots s_l|m;r_1,\ldots,r_k\rangle 
=\delta_{k,l}\,\delta_{m,n}
\sum_{\pi\in S_k}
\prod_{i=1}^k\, \delta(s_i-r_{\pi(i)})
\label{newinnerp}
\end{equation}
This will be used only as a tool to isolate 
certain matrix elements of the Hamiltonian
by the following representation of the unit operator
\begin{equation}
1=\sum_{k=0}^\infty\frac{1}{k\,!}\,
\int\limits_{0\leq r_0,r_1,\ldots, r_k\leq 1}\!\!\!\! d^k r\,\sum_{n=-\infty}^\infty 
 |n;r_1,\ldots,r_k\rangle  \langle n;r_1,\ldots,r_k|.
\end{equation}
It enables us to write compact expressions 
for the corrections to the eigenvalues. 
With this scalar product our Hamiltonian is not Hermitian. The scaling
dimensions, however, have to be real and it is not hard to see that they
are: Letting $H$ act on $\bar{\cal O}_{\bar\beta}$ instead
of ${\cal O}_{\alpha}$ in (8) we get $T=S H^{\dagger}$ and thus
$H=S H^{\dagger} S^{-1}$ which implies that
$H=\lambda/4\pi^2 \cdot h$ has real eigenvalues.
Clearly, the eigenvalues and eigenvectors of $h$
are independent of the definition of a scalar product.
Using the scalar product and eq.~\eqref{h-on-eigen} 
one easily writes down the matrix elements of the interaction 
$h'=h_+ + h_-$ in the momentum basis, for instance
\begin{myeqnarray}\label{matrixelements}
\langle m;r| h' |n\rangle \eq
-\frac{1}{(1-r)^{3/2}}\, \frac{4m}{n-\frac{m}{1-r}}
\, \sin^2(\pi n r), \nln
\langle n| h' |m;r \rangle\eq\frac{r}{(1-r)^{1/2}}
\, \frac{4n }{n-\frac{m}{1-r}}\,
\sin^2(\pi n r). 
\end{myeqnarray}

The corrections to the energy of a general state
$|\alpha\rangle$ are then given by standard formulas 
of quantum mechanical perturbation theory. The first
two non-vanishing orders of non-degenerate perturbation theory give
\begin{myeqnarray} \label{E2}
E^{(1)}_{|\alpha\rangle} \langle \alpha|\alpha\rangle\eq \langle \alpha|h'\, 
\Delta_{|\alpha\rangle}\, h' | \alpha\rangle,
\nln
E^{(2)}_{|\alpha\rangle} \langle \alpha|\alpha\rangle\eq 
\langle \alpha|h'    \,\Delta_{|\alpha\rangle}\,h'
                     \,\Delta_{|\alpha\rangle}\,h'
                     \, \Delta_{|\alpha\rangle}\,h' |\alpha\rangle 
\nl
\qquad - E^{(1)}_{|\alpha\rangle}\, \langle \alpha|h' \,
(\Delta_{|\alpha\rangle})^2\,h'|\alpha\rangle.
\end{myeqnarray}
Here the free propagator $\Delta_{|\alpha\rangle}$ is defined as
\begin{equation}
\Delta_{|\alpha\rangle}= \frac{1-|\alpha\rangle\langle \alpha|}{E^{(0)}_{|\alpha\rangle}-h_0}.
\end{equation}
Some technical complications potentially
arise due to the fact that the free spectrum eq.\eqref{freespectrum} 
is degenerate 
as was first noticed in \cite{Constable:2002vq}
(\emph{cf.} discussion after eq.\eqref{1to3}).

We now calculate the energy shift at genus one and two.
Determining the energy shift at genus one 
is a one-line computation
\begin{equation}
E^{(1)}_{|n\rangle}
=\int\limits_0^1 dr  \sum_{m=-\infty}^\infty \langle n|h'|m;r\rangle \, 
\frac{1}{4\pi^2\bigbrk{n^2-\frac{m^2}{(1-r)^2}}}\, \langle m;r|h'|n\rangle
= \frac{1}{12}+\frac{35}{32\,\pi^2n^2}
\label{oneline}
\end{equation}
and implies a correction to the anomalous dimension 
in agreement with the result of 
references~\cite{Beisert:2002bb,Constable:2002vq}.
This computation was originally performed in \cite{Janik:2002bd},
however, it relied on the assumption 
that there are no $\order{g_2^2}$ contact-terms in $h$;
this fact is manifest in our formalism.

Having convinced ourselves of the simplicity
of this method we proceed to genus two,
\emph{cf.} eq.\eqref{E2}.
Using~\eqref{matrixelements} one readily computes
\begin{equation}
\langle n|h' \,(\Delta_{|n\rangle})^2\, h'|n\rangle= 
-\frac{1}{120}+\frac{1}{32\,\pi^2n^2}
-\frac{21}{256\,\pi^4 n^4}.
\end{equation}
The computation of the first term of~\eqref{E2}
splits into two parts, namely a contribution from a single-trace
channel and a contribution from a triple-trace channel (corresponding
to the number of traces of the BMN operators changing as
1-2-1-2-1 and 1-2-3-2-1 respectively).

For the evaluation of the contribution from 
the single-trace channel term one needs the
matrix elements:
\begin{myeqnarray}
\langle l|h'\, \Delta_{|n \rangle}\, h'|n\rangle \eq
\frac{l}{n}\,\langle n|h'\, \Delta_{|n\rangle}\, h'|l\rangle, \nln
\langle n|h'\, \Delta_{|n\rangle}\, h'|l\rangle\eq
\frac{n}{6(n-l)}+ \frac{l^6-l^4\, n^2+6\,l^2\, n^4 - 2n^6}{4\pi^2\,n\,(l-n)^3\,l^2\,(n+l)^2}.
\end{myeqnarray}
Using these yields
\begin{equation}E^{(2),\mathrm{single}}_{|n\rangle}
=-\frac{1}{2688} 
+ \frac{107}{9216\,\pi^2n^2}
- \frac{695}{12288\,\pi^4n^4} 
- \frac{3785}{8192\,\pi^6n^6}.\end{equation}
Turning to the final computation of the triple-trace channel contribution we
note the intermediate formulas
\begin{myeqnarray}\label{1to3} 
\langle l;s_1,s_2| h'\Delta_{|n\rangle}\, h'|n\rangle \eq
\frac{l}{n(1-s_1-s_2) s_1 s_2}\, 
\langle n| h' \Delta_{|n\rangle}\, h'|l;s_1,s_2\rangle ,
\\
\langle n| h' \Delta_{|n\rangle}\, h'|l;s_1,s_2\rangle \eq
-\frac{8}{\pi^2}\, \frac{s_1\, s_2\, (1-s_2)^3}{\sqrt{1-s_1-s_2}}  
\,n \, \sin^2(\pi n s_2)\, \left(n+\frac{l}{1-s_1-s_2}\right) \nl
\quad\times\sum_{m=1}^\infty \frac{m^2\, \sin^2(\pi\, m\, \frac{s_1}{1-s_2})}
{\Bigl(m^2-n^2 
(1-s_2)^2\Bigr)^2\, \Bigl(m^2-l^2\,
(\frac{1-s_2}{1-s_1-s_2})^2\Bigr)}
\nl
 + (s_1\leftrightarrow s_2). \nn
\end{myeqnarray}
The single-trace state $|n\rangle$ and the triple-trace state $|l;s_1,s_2\rangle$ are
degenerate if $n=\pm \frac{l}{(1-s_1-s_2)}$ 
which can happen only for $n>1$. The
matrix element~\eqref{1to3} is easily seen to vanish for 
$n=-\frac{l}{1-s_1-s_2}$. However for $n=\frac{l}{1-s_1-s_2}$ it is 
finite
For $n>1$ this observation casts some doubt on the use of 
non-degenerate perturbation theory; however, 
for $n=1$ there clearly is no problem \cite{Constable:2002vq}.
The above sum can be 
performed with the help of the formula
\begin{equation}
\sum_{k=1}^\infty\frac{\sin^2(\pi\, k\, x)}{k^2-b^2}=
\frac{\pi}{2b}\, \frac{\sin(\pi b x)\, \sin (\pi b (1-x))}{\sin(\pi b)}
\end{equation}
for $0\leq x\leq 1$ and  $b$ non-integer. After some 
algebra using \texttt{Mathematica} one obtains
\begin{myeqnarray}
E^{(2),\mathrm{triple}}_{|n\rangle}
\eq
\frac{1}{2}
\sum_{l=-\infty}^\infty\,\,\, \int\limits_{0\leq s_1+s_2\leq 1}\!\!\!\!\! ds_1\, ds_2\,
\frac{\langle n| h' \Delta_{|n\rangle}\, h'|l;s_1,s_2\rangle
\, 
\langle l;s_1,s_2| h' \Delta_{|n\rangle}\, h'|n\rangle}
{4\pi^2\bigbrk{n^2-\frac{l^2}{(1-s_1-s_2)^2}}}\, 
\nle
 -\frac{13}{40320} 
-\frac{47}{2560\,\pi^2n^2} 
+ \frac{97}{768\,\pi^4n^4} 
+ \frac{385}{16384\,\pi^6n^6}
+\int_{0}^1 ds_0\, f_n(s_0).
\nln
\end{myeqnarray}
The function that remains to be integrated is
\begin{equation}f_n(s_0)=-\frac{(1-s_0)^2 s_0(15+4\pi^2 s_0^2)\cot(\pi n s_0)}{256\,\pi^3 n^3}.\end{equation}
For $n=1$ the integral is finite and easily evaluated to be
\begin{equation}
\int_0^1 dr_0\,f_1(r_0)=
-\frac{\zeta(3)}{128\,\pi^4 }
-\frac{45\,\zeta(3)}{512\,\pi^6 }
+\frac{15\,\zeta(5)}{128\,\pi^6}
\label{integral}
\end{equation}
For $n>1$ there are poles at 
$s_0=\sfrac{m}{n}$, $1\leq m\leq n-1$. These are
related to the above discussed degeneracy of the single-trace 
state $|n\rangle$ with the triple-trace states
$|m;s_1,s_2\rangle$ where $s_0=1-s_1-s_2=\sfrac{m}{n}$.
Deferring this problem to future work, we proceed by regulating
the integral by a principle value prescription. We find 
the same result as eq.\eqref{integral} except for the replacement
$\pi \rightarrow \pi n$.
In total we get for the one-loop two-torus contribution
\begin{myeqnarray}
E^{(2)}_{|n\rangle}\eq-\frac{11}{46080}\,\,\frac{1}{\pi^2n^2 }
+\left(\frac{521}{12288}-\frac{\zeta(3)}{128}\right)\frac{1}{\pi^4 n^4}
\nl\qquad
+
\left(-\frac{5715}{16384} - \frac{45\,\zeta(3)}{512} +
  \frac{15\,\zeta(5)}{128}
\right)\frac{1}{\pi^6n^6}.
\end{myeqnarray}
Our derivation of $E^{(2)}_{|n\rangle}$ is rigorous only for $n=1$;
it would be important to more carefully examine the validity of 
this formula for $n>1$.

As we discussed in the beginning, the various terms {\it do not
cancel}, and thus $E^{(2)}_{|n\rangle} \neq 0$,
disproving the idea that the genus counting parameter
$g_2^2$ should always be accompanied by a factor of $\lambda'$.
However, interestingly, the $n$-independent, constant terms do cancel. 
This suggests, together with the result of the sphere and the torus,
that the exact energy might scale as
\begin{equation}\label{eq:conjecture}
E_{|n \rangle} \sim 4 \pi^2 n^2
\sum_{h=0}^\infty \,c_h \left( \frac{\gtwo^2}{4 \pi^2 n^2} \right)^h
\quad {\rm for} \quad n \rightarrow \infty,
\end{equation}
where the $c_h$ are numerical constants: 
$c_0=1, c_1=\frac{1}{12}, c_2=-\frac{11}{6! 2^4}, \ldots$.
 
We can also find the scaling dimensions of multi-trace operators
from our formalism.
The result \eqref{speculation} was first derived, 
using different methods, in \cite{Gursoy:2002fj}.
Some extra care has to be taken due to the
continuous spectrum of multi-trace operators, {\it cf.} 
eq.\eqref{freespectrum}. In particular, multi-trace
operators are delta-function normalized, see eq.\eqref{newinnerp}.
A straightforward repetition of perturbation theory,
as in eq.\eqref{E2}, for multi-trace operators 
reveals that the relevant contribution,
proportional to the normalizing delta function, results solely 
from the disconnected channels. The latter are defined as the contributions
where the external traces without impurities do not 
participate in the interaction.
Assuming that no subtleties arise from divergences of
connected channels (as we have explicitly verified for genus one)
we find
\begin{equation}
E^{(h)}_{|n;r_1,\ldots,r_k\rangle}=r_0^{4h-2} E^{(h)}_{|n\rangle}.
\label{speculation}
\end{equation}
It would be important to more carefully investigate this issue
in the framework of our formalism.

Clearly it should be interesting to extract further information from
our Hamiltonian formulation. In particular, it would be exciting
to solve the eigenvalue problem in the WKB limit, as this might lead to 
non-perturbative insights into the genus expansion, and, in
consequence, to non-perturbative results on plane-wave strings,
{\it cf.} eq.\eqref{eq:conjecture}.
It will also be very interesting to understand the terms one needs 
to add to the Hamiltonian in order to include the effects of
radiative corrections beyond one loop, and of more than two
impurities. We do not see any reason why this should not be
possible, and therefore conjecture that BMN gauge theory can quite
generally be reformulated as a quantum mechanical system.

Many of the above methods and insights are already present in the
more difficult case of finite $J$. The novel ``BMN'' way of looking
at ${\cal N}=4$ gauge theory is beginning to lead to startling
progress beyond the admittedly somewhat artificial limit of 
infinitely large $R$-charges. In particular, in
\cite{Beisert:2002tn} it was shown that finite $J$ versions of
BMN operators can be rigorously defined, and should be
considered to be generalizations of the
simplest unprotected field in ${\cal N}=4$, the so-called Konishi 
scalar. They are the 
``next to best thing to BPS operators'', as they are,
in a rather precise sense, ``almost protected''.
This discovery has lead to
a fresh look at the representation theory of ${\cal N}=4$ operators, and 
to explicit results for anomalous dimensions of whole families
of operators \cite{Beisert:2002tn}. In this context we
should mention a very interesting, recent paper by Minahan and
Zarembo \cite{Minahan:2002ve}. 
In a spirit conceptually very close to the present work
they are considering the planar, one-loop diagonalization of
more complicated operators with many impurities, uncovering
some of the structures (e.g.~Bethe-Ansatz) found in integrable spin chains. 
One could therefore hope that currently known results are just
the beginning of the discovery of an integrable sector in
${\cal N}=4$ gauge theory. As is frequently the case for two-dimensional 
integrable systems, such structures might exist both on the 
lattice (here: finite $J$) and in the continuum limit 
(here: $J \rightarrow \infty$).

\vspace{6mm}
\noindent
{\bf Acknowledgments} 

\noindent
We would like to thank Stefano Kovacs for interesting 
discussions. N.B.~dankt der \emph{Studienstiftung des
deutschen Volkes} f\"ur die Unterst\"utzung durch ein 
Promotions\-f\"orderungsstipendium.

\end{document}